\documentclass[10pt,sigplan,letterpaper,twocolumn]{article}
\usepackage[10pt,nocopyright]{sigmin}

\usepackage{listings}
\usepackage{color}
\usepackage{epsfig,makecell}    
\usepackage{ifpdf}
\ifpdf
  \usepackage{epstopdf}
\fi                                  
\usepackage[square,comma,numbers,sort&compress]{natbib}          
\usepackage[hyphens]{url}                                         
\usepackage[breaklinks,colorlinks]{hyperref}                      
\usepackage[usenames,dvipsnames]{xcolor}                          
\hypersetup{citecolor=blue,linkcolor=blue}                        
\usepackage{amsmath,amsopn,amssymb,amsthm}                        
\usepackage{thmtools}
\usepackage[toc,page]{appendix}
\usepackage{hyperref}
\usepackage{subfigure}
\usepackage{endnotes,microtype,xspace,fancyvrb,multirow} 
\usepackage{booktabs}                                             
\usepackage{array,underscore,relsize}                             
\usepackage[T1]{fontenc}                                          
\usepackage{times}                                                
\usepackage{fancyhdr,lastpage}                                    
\usepackage{enumitem}                                             
\usepackage[labelfont=bf,font=normal,skip=1pt]{caption}           
\usepackage[export]{adjustbox}                                    
\usepackage{breakurl}                                             
\usepackage{pifont}                                               
\usepackage{tablefootnote}                                        
\usepackage{graphicx}
\usepackage[linesnumbered,vlined,ruled,commentsnumbered]{algorithm2e}
\DontPrintSemicolon
\newcommand{\ra}[1]{\renewcommand{\arraystretch}{#1}}
\SetCommentSty{mycommfont}

\usepackage{amssymb}
\usepackage[normalem]{ulem}
\usepackage[hang,flushmargin]{footmisc}
\usepackage{textcomp}
\usepackage{graphicx}

\usepackage[compact]{titlesec}
\titleformat*{\section}{\large\bfseries}
\titleformat*{\subsection}{\normalsize\bfseries}
\titleformat*{\subsubsection}{\normalsize\bfseries}
\titlespacing*{\section}{0pt}{*2}{*1}
\titlespacing*{\subsection}{0pt}{*1.5}{*0.8}

\definecolor{mygreen}{rgb}{0,0.6,0}  
\definecolor{mygray}{rgb}{0.5,0.5,0.5}  
\definecolor{mymauve}{rgb}{0.58,0,0.82}

\hypersetup{
  colorlinks,
  linkcolor={red!50!black},
  citecolor={blue!50!black},
  urlcolor={blue!80!black}
}

\newcommand{\sys}{{\textsc{XAuto}}}
\newcommand{\xnode}{{\textsc{XNode}}}
\newcommand{\xnodes}{{\textsc{XNode}s}}
\newcommand{\xmod}{{\textsc{XModule}}}
\newcommand{\xmods}{{\textsc{XModule}s}}

\newcommand{\sysH}{{{\sys}-H}}

\newcommand{\us}{{{$\upmu$}s}}

\newcommand{\stitle}[1]{\vspace{1.ex}\noindent{\bf #1}}
\newcommand{\nostitle}[1]{\noindent{\bf #1}}
\newcommand{\etitle}[1]{\vspace{0.8ex}\noindent{\em\underline{#1}}}

\newcommand{\code}[1]{{\smaller[0.9]{{\texttt{#1}}}}}

\newcommand{\squishlist}{
  \begin{list}{$\bullet$}{
    \setlength{\itemsep}{0pt}
    \setlength{\parsep}{3pt}
    \setlength{\topsep}{3pt}
    \setlength{\partopsep}{0pt}
    \setlength{\leftmargin}{1.5em}
    \setlength{\labelwidth}{1em}
    \setlength{\labelsep}{0.5em}
  }
}

\newcommand{\squishend}{
  \end{list}
}

\usepackage{tikz}

\usepackage{balance}
\usepackage{upgreek}

\begin{document}

\title{\LARGE \bf Holistic Heterogeneous Scheduling for Autonomous \\ Applications using Fine-grained, Multi-XPU Abstraction}

\author{
    Mingcong Han,\;
    Weihang Shen,\;
    Rong Chen\thanks{Rong Chen is the corresponding author (\url{rongchen@sjtu.edu.cn}).},\;
    Binyu Zang
    and\; Haibo Chen\;\\[5pt]
    \normalsize{Institute of Parallel and Distributed Systems, Shanghai Jiao Tong University}
}

\date{}
\maketitle

\frenchspacing

\begin{abstract}
Modern autonomous applications are increasingly utilizing
multiple heterogeneous processors (XPUs) to accelerate
different stages of algorithm modules.
However, existing runtime systems for these applications,
such as ROS, can only perform module-level task management,
lacking awareness of the fine-grained usage of multiple XPUs.
This paper presents {\sys}, a runtime system designed to
cooperatively manage XPUs for latency-sensitive autonomous applications. 
The key idea is a fine-grained, multi-XPU 
programming abstraction---{\xnode}, which aligns with the stage-level 
task granularity and can accommodate multiple XPU implementations.
{\sys} holistically assigns XPUs to {\xnodes} and 
schedules their execution to minimize end-to-end latency.
Experimental results show that {\sys} can reduce the end-to-end latency
of a perception pipeline for autonomous driving by 1.61$\times$ 
compared to a state-of-the-art module-level scheduling system (ROS2).
\end{abstract}

\section{Introduction}

Autonomous things (abbreviated as AuT), including autonomous
vehicles~\cite{taeihaghGoverningAutonomousVehicles2019,thrunRoboticCars2010},
drones~\cite{floreanoScienceTechnologyFuture2015},
and robots~\cite{fahimiAutonomousRobotsModeling2009,ingrandDeliberationAutonomousRobots2017},
are becoming increasingly popular in applications, such as
transportation~\cite{bagloeeAutonomousVehiclesChallenges2016,pisarovUseAutonomousVehicles2021},
delivery~\cite{schlentherPotentialPrivateAutonomous2020},
and surveillance~\cite{ayubNextGenerationSecurity2018}.

The rapid growth of AuT largely stems from advancements 
in artificial intelligence (AI) algorithms.
These algorithms empower them to perceive their environments, make decisions, and take actions 
as effectively, if not more so, than humans.

\begin{figure}[t]
    \begin{minipage}{1\linewidth}
		\centering\includegraphics[width=1.\linewidth]{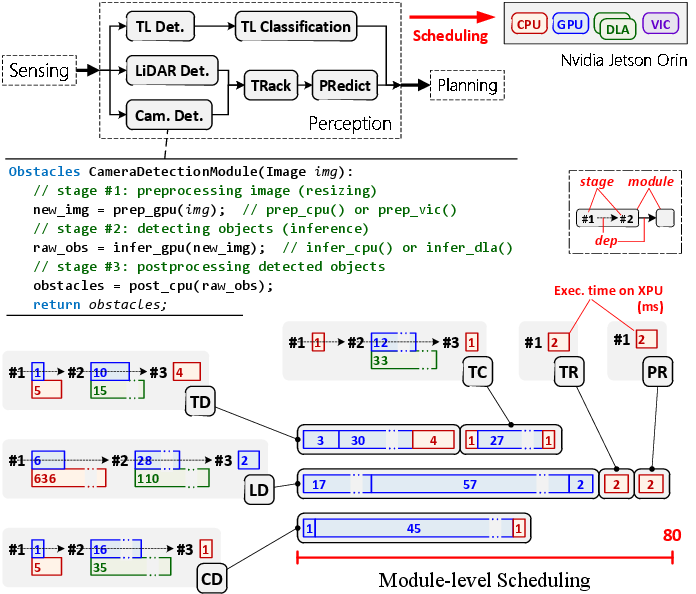} 
    \end{minipage} \\ [8pt]
    \begin{minipage}{1\linewidth}
    \caption{\emph{\small{An example of a ROS-like module-level scheduling system
    scheduling the perception component~\cite{ApolloPerception} of Autoware.Auto~\cite{AutowareGithub}
    on NVIDIA Jetson AGX Orin~\cite{NVIDIAOrin}.
    \textbf{Top Left}: The modules contained in the perception component and
    their dependencies. The gray box represents a module (e.g., TL Det.), and the solid arrows
    represent the dependencies between modules.
    \textbf{Middle Left}: Pseudocode of the camera detection module.
    The functions commented out behind each stage indicate the alternative implementations.
    \textbf{Bottom Left}: The stages contained in each module and their execution
    times on different XPUs. Here, \#n represents the n-th stage in the module, and
    the numbers in the colored boxes represent the execution time of this
    stage on the corresponding XPU (e.g., blue for GPU).
    The dashed arrows represent the dependencies between stages.
    \textbf{Bottom Right}: The timeline of using module-level scheduling.}}}
    \label{fig:motiv-example}
    \end{minipage} \\[-10pt]
\end{figure}

AuT applications typically comprise tens of algorithms orchestrated to achieve
the overall goal.
For example, the perception component of Autoware.Auto autonomous driving
system~\cite{AutowareGithub} consists of six algorithm modules, which have dependencies and operate
concurrently with each other shown in Figure~\ref{fig:motiv-example}. 
This inherent complexity has led to a modular development approach, where each module
implements a specific algorithmic function.
In these systems, end-to-end latency---defined as the duration from sensor data acquisition
to the decision making---represents the most critical performance metric, as it directly
impacts the system safety and reliability~\cite{Lin2018TheAI,Yu2020BuildingTC,Hadidi2021QuantifyingTD}.
To illustrate this criticality, consider that a mere 100\,ms computation delay can reduce an autonomous
vehicle's obstacle avoidance range by 3.5\,meters~\cite{Yu2020BuildingTC}.

The computational demands of these algorithms necessitate deployment of AuT applications on 
heterogeneous computing platforms~\cite{Sung2021EnablingLA,Kato2018AutowareOB,Talpes2020ComputeSF,Yu2020BuildingTC}
that are equipped with multiple types of processing units (XPUs),
such as NVIDIA Jetson series~\cite{NVIDIAXavier,NVIDIAOrin}, Tesla FSD Hardware series~\cite{Talpes2020ComputeSF}, 
and AMD Versal AI Edge series~\cite{amdversalai}. 
A representative example is the NVIDIA Jetson AGX Orin~\cite{NVIDIAOrin}, which combines 
a CPU, a GPU, two Deep Learning Accelerators (DLAs, specialized NPUs), two Video Image 
Compositors (VICs), and one Programmable Vision Accelerator (PVA).
Each XPU type exhibits distinct characteristics in terms of functionality 
and performance. For instance, GPUs excel at parallel computing tasks, while NPUs 
target deep learning inference tasks.

Due to hardware specialization, ASICs like DLAs are designed to perform specific 
computational functions, limiting their applicability to particular \textit{stages} within 
an algorithm module. Consider the camera detection module illustrated in 
Figure~\ref{fig:motiv-example}, which comprises three distinct stages: preprocessing, 
model inference, and postprocessing. Each stage can be executed on different XPUs based 
on their capabilities---preprocessing can utilize CPU, GPU, or VIC, while model inference 
can be performed on CPU, GPU, or DLA.

Despite the widespread adoption of heterogeneous platforms, 
current runtime systems for AuT applications---including ROS~\cite{ROSPaper,ROS2Paper}, CyberRT~\cite{CyberRT}, and ERDOS~\cite{Gog2022D3AD}---lack XPU management capabilities, 
ultimately resulting in inefficient XPU usage and high end-to-end latency.
This inefficiency stems from two primary limitations.
First, these systems are restricted to scheduling CPU threads, delegating the scheduling
of other XPUs, including GPUs and NPUs, to the underlying hardware schedulers.
These hardware schedulers, however, operate without knowledge of inter-module dependencies
and concurrency, leading to inefficient execution order.
Second, these systems delegate XPU assignment decisions (i.e., which XPU to use for each 
algorithm module) to developers rather than managing them systematically.
Without a global perspective of XPU utilization across
modules, developers tend to make locally optimal but globally suboptimal choices,
often defaulting to the fastest available XPU for each module. This approach
frequently results in resource contention and degraded overall performance.

The fundamental limitation lies in the programming abstractions of current systems.
In frameworks like ROS and CyberRT, the basic unit of abstraction (a ROS node or
CyberRT component) represents an entire algorithm module as a single CPU function.
This coarse-grained (\textit{module-level}) single-XPU abstraction is inadequate  
to represent the fine-grained (\textit{stage-level}) multi-XPU usage of AuT applications.

To this end, we propose {\xnode}, a fine-grained multi-XPU abstraction for programming
AuT applications on heterogeneous computing platforms.
Unlike the coarse-grained modules that are divided based on algorithmic functionality,
an {\xnode} represents a specific stage within an algorithm, divided according to the 
XPU usage. An {\xnode} can have multiple implementations on different XPUs, 
with one implementation for one XPU.

This abstraction allows developers to 
concentrate on XPU-specific implementations, abstracting away the 
complexities of XPU selection and inter-stage coordination. Furthermore, it enables the 
runtime system to make informed decisions about XPU assignment and scheduling by leveraging 
comprehensive knowledge of each {\xnode}'s XPU requirements and inter-node dependencies.

Based on the {\xnode} abstraction, we present {\sys}, a framework for 
deploying and scheduling AuT applications on heterogeneous computing platforms.
{\sys} leverages the {\xnode} abstraction through two key mechanisms:
(1) holistic XPU assignment, which automatically optimizes the XPU selection 
for all {\xnodes} simultaneously during deployment, considering their interactions
and resource contention, and
(2) fine-grained XPU scheduling, which enables independent scheduling of tasks 
at stage boundaries across different XPUs, allowing for precise control over 
execution order.

The flexibility provided by the {\xnode} abstraction introduces significant 
challenges for designing an optimal scheduling policy due to its expanded solution space. 
An application consists of multiple modules, each containing multiple {\xnodes}, and each {\xnode} 
has multiple XPU implementations.
To illustrate the complexity, consider that a system with 20 {\xnodes} and 3 XPU implementations,
which yields approximately $10^{27}$ ($3^{20}{\times}20!)$ possible combinations 
of assignments and priorities.
To efficiently navigate this tremendous solution space, {\sys} employs an Integer Linear
Programming (ILP) model to determine XPU assignments and scheduling parameters.
This model optimizes end-to-end latency by considering both the {\xnode} topology
and the execution time profiles of various XPU implementations.

We adapted {\sys} for the NVIDIA Jetson AGX Orin platform and implemented
priority-based preemptive scheduling for the CPU, GPU, and DLA.
We evaluate the performance of {\sys} using synthetic workloads with varying numbers 
of {\xnodes}, and compare it with ROS2~\cite{ROS2Paper} (a state-of-the-art module-level system).
Due to the lack of automatic XPU assignment capability and the lack of a global view 
of XPU utilization in ROS2, we manually pre-assign XPUs to {\xnodes} in ROS2, 
using a local-optimal policy that assigns the fastest available XPU to each {\xnode}.
The experimental results show that {\sys} consistently outperforms 
ROS2, resulting in end-to-end latency improvements 
ranging from 1.23$\times$ to 2.01$\times$.
Additionally, we conducted an application-level case study based on an autonomous driving 
perception component, where {\sys} achieved 1.61$\times$ performance improvement 
compared to ROS2. 
Even with the same assignment policy, {\sys} still outperforms ROS2 by up to 1.29$\times$.
We plan to open-source {\sys} and all associated workloads.

\stitle{Contributions.} We summarize our contributions as follows:
\begin{itemize}

    \item A fine-grained multi-XPU abstraction, {\xnode}, that enables stage-level XPU-aware 
    task management for AuT applications. ($\S$\ref{sec:prog})
    
    \item A runtime system, {\sys}, that deploys and schedules {\xnodes} on 
    heterogeneous computing platforms with priority-based scheduling support. 
	($\S$\ref{sec:sys} and $\S$\ref{sec:sched})
    
    \item An integer linear programming (ILP) model that facilitates efficient decision-making 
	regarding XPU assignment and scheduling problems for {\xnodes}. ($\S$\ref{sec:policy})
    
    \item An experimental evaluation that demonstrates the efficacy and efficiency of {\sys}. ($\S$\ref{sec:eval})
\end{itemize}

\section{Background and Motivation}\label{sec:bg}

\begin{figure}[t]
    \vspace{-2mm}
    
    \begin{minipage}{1\linewidth}
        \centering\includegraphics[width=\linewidth]{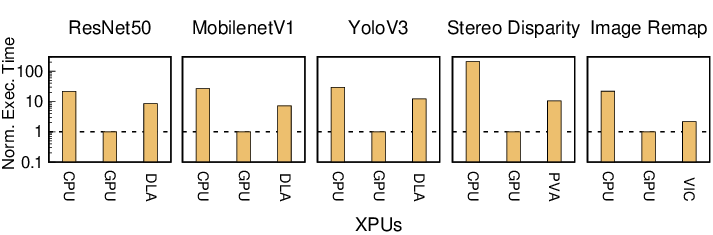} 
    \end{minipage} \\[5pt]

    \begin{minipage}{1\linewidth}
        \caption{\emph{\small{
        The execution time of five algorithms on supported XPUs including CPU,
        GPU, DLA, VIC, and PVA, normalized with the results on GPU.
        ResNet50~\cite{He2016DeepRL}, MobilenetV1~\cite{Howard2017MobileNetsEC}, and YoloV3~\cite{redmon2018yolov3}
        are executed using ONNXRuntime for CPU, and TensorRT for GPU and DLA.
        The others are executed using the NVIDIA VPI library.}}}
    \label{fig:motiv-tradeoff}
    \end{minipage} \\[-10pt]
\end{figure}

\subsection{Characterizing Heterogeneous Processors}
\label{sec:bg:xpu}

AuT applications are commonly deployed on computing systems with heterogeneous processors (XPUs),
such as CPUs, GPUs, NPUs, and ASICs~\cite{NVIDIAXavier,NVIDIAOrin,Talpes2020ComputeSF,amdversalai,QualcommSnapdragon}.
We take NVIDIA Jetson AGX Orin~\cite{NVIDIAOrin} as an example, which is the most popular SoC for
AuT applications and has been widely adopted by the industry.
The SoC integrates multiple processors: a 12-core ARM CPU, a GPU, two Deep 
Learning Accelerators~\cite{NvidiaDLA} (DLAs),
one Programmable Vision Accelerator (PVA), and two Vision Image Compositors (VICs).
These XPUs offer distinct characteristics summarized as follows:

\stitle{Functionality.}
Different XPUs have different functionalities and are suitable for executing different types of algorithms.
For example, GPUs are suitable for executing data-parallel algorithms, and DLAs are only capable of executing
deep-learning model inferences.
PVA is designed for executing computer vision algorithms, such as image filtering and feature detection.
VIC can be used to accelerate image processing tasks, such as image scaling and remapping.
Hardware vendors provide libraries that implement algorithms on different XPUs.
For example, NVIDIA TensorRT~\cite{TensorRT} allows a DL model to run either on GPU or DLA with only one line of code change,
while NVIDIA VPI~\cite{NvidiaVPI} provides a set of APIs for computer vision algorithms with implementations on various XPUs
including CPU, GPU, PVA and VIC.

\stitle{Performance.}
While various XPUs can execute similar algorithms, they exhibit distinct performance characteristics.
Figure~\ref{fig:motiv-tradeoff} demonstrates the execution time variations across five algorithms on different XPUs.
The performance hierarchy typically shows GPUs achieving the highest performance, followed by specialized ASICs
(including DLA, PVA, and VIC), with CPUs demonstrating comparatively lower performance.
This performance distribution creates a dual effect: developers naturally gravitate toward GPU implementations
to optimize performance, yet this preference leads to increased GPU contention, potentially becoming
a bottleneck in overall performance.

\subsection{Characterizing AuT Applications}
\label{sec:bg:ads}

\nostitle{Latency-sensitive.} 
AuT applications are highly sensitive to the end-to-end latency from getting sensor data 
to making decisions, which significantly impacts the safety and reliability of these systems~\cite{Yu2020BuildingTC,goodall2017rise,Lin2018TheAI,Murray2016TheMO,Boroujerdian2018MAVBenchMA}.
For example, autonomous vehicles must make decisions within 100\,ms
to prevent accidents~\cite{Lin2018TheAI}.
A delay of just 30\,ms will reduce the obstacle avoidance range for vehicles 
on the highway by 1\,meter~\cite{Yu2020BuildingTC}, which dramatically
increases the risk of collisions~\cite{ReactTime, ReactTime2}.
Similarly, drones have to achieve millisecond-scale response latency to
support high-speed flights~\cite{Boroujerdian2018MAVBenchMA}.

\stitle{Multitask-cooperative.} 
AuT applications typically consist of multiple algorithm modules having
\emph{dependency} and \emph{concurrency} relationships. 
In autonomous driving, for example, the perception module detects 
objects, lanes, and traffic signs from sensor data simultaneously, while 
the planning module uses these outputs to plan driving paths.
As the complexity of AuT applications has grown,
the number of tasks involved has also increased significantly. 
For example, Autoware~\cite{AutowareGithub} includes more than 30 tasks.
To simplify the development and maintenance of these applications,
developers typically adopt a \textit{modular design}, 
with each module responsible for handling a specific algorithm.

\stitle{Stage-partitioned.}
To meet stringent latency requirements, algorithms in AuT applications commonly rely on 
heterogeneous XPUs for acceleration. However, due to the specialized nature of XPUs, 
a single XPU often lacks the capability to execute an entire algorithm independently. 
Consequently, algorithms must be decomposed into multiple stages, with each stage 
optimized for execution on a specific XPU. For example, as illustrated in 
Figure~\ref{fig:motiv-example}, a camera-based obstacle detection algorithm comprises 
three distinct stages: preprocessing (executed on CPU or VIC), neural network inference 
(executed on GPU or DLA), and postprocessing (executed on CPU).

\begin{figure}[t]
    \begin{minipage}{1\linewidth}
        \centering\includegraphics[width=1.02\linewidth]{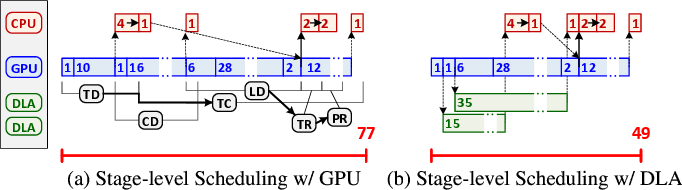} 
    \end{minipage} \\[5pt]
    \begin{minipage}{1\linewidth}
    \caption{\emph{\small{The timeline of using stage-level scheduling on the
    same perception component as in Figure~\ref{fig:motiv-example}.
    \textbf{(a)} Stage-level scheduling with GPU-only model inferences.
    \textbf{(b)} Stage-level scheduling with DLA-enabled model inferences.
    }}}
    \label{fig:xdag-example}
    \end{minipage} \\[-10pt]
\end{figure}

\subsection{Limitation of Existing Approaches}
\label{sec:bg:motiv}

Existing frameworks used for deploying AuT applications, such as ROS~\cite{ROSPaper,ROS2Paper}, 
CyberRT~\cite{CyberRT}, and ERDOS~\cite{Gog2022D3AD}, struggle to effectively manage XPU resources, 
significantly hampering their performance on modern heterogeneous platforms. The performance issues 
primarily stem from two factors: XPU scheduling and XPU assignment. We demonstrate these limitations 
using the perception component of Autoware.Auto~\cite{AutowareGithub} deployed on NVIDIA Jetson AGX Orin, 
as shown in Figure~\ref{fig:motiv-example}.

\etitle{XPU scheduling.}
Existing frameworks only schedule CPU tasks, leaving other XPU scheduling to hardware
and drivers that use simple policies like round-robin. 
However, a smarter scheduling policy that considers XPU usage and task dependencies 
could potentially reduce latency. For example, in Figure~\ref{fig:motiv-example}, 
while \textit{TR} and \textit{PR} modules (CPU-only) appear at the end of the pipeline, 
they could run in parallel with \textit{TC}'s GPU stage, reducing latency as shown in 
Figure~\ref{fig:xdag-example}(a). This requires prioritizing GPU stages in \textit{LD} and 
\textit{CD} over \textit{TC}, while letting \textit{TR} and \textit{PR} start when their 
dependencies are met. This capability is however missing in current frameworks.

\etitle{XPU assignment.}
Existing systems are unable to dynamically determine the optimal assignments of XPUs 
to algorithms. For example, Autoware.Auto requires developers to manually configure 
the use of DLAs~\cite{AutowareDLAConfig}. 
Without a system-wide view of XPU utilization, developers often make locally-optimal but 
globally-suboptimal decisions (e.g., defaulting to GPU for all model inference tasks).
While GPUs offer superior performance, defaulting to GPU can cause GPU congestion
and increase end-to-end latency. As shown in Figure~\ref{fig:xdag-example}(b), distributing 
tasks across slower but idle XPUs (like DLAs) can significantly reduce latency compared to GPU-only execution.

The fundamental issue is that existing systems are constrained by 
their scheduling abstractions.
Their basic execution units---whether ROS nodes or 
CyberRT components---are abstracted as CPU-only functions
operating at the granularity of algorithm modules,
which cannot reflect the multi-XPU usage inside the modules,
and misaligns with the actual granularity of XPU usage.

\stitle{Key idea: fine-grained, multi-XPU abstraction.}
We argue that the systems for AuT applications should natively support fine-grained (stage-level) 
multi-XPU abstraction. This abstraction provides two key benefits:
First, it enables precise task scheduling at stage boundaries, offering greater flexibility 
to make XPU scheduling decisions. 
For example, as shown in Figure~\ref{fig:xdag-example}(a), the system can prioritize critical GPU 
stages in \textit{LD} and \textit{CD} over \textit{TC}, reducing end-to-end latency.
Second, it enables holistic XPU assignment by providing a system-wide view of 
stage requirements. As demonstrated in Figure~\ref{fig:xdag-example}(b), this abstraction allows
optimal resource allocation --- assigning \textit{TD} and \textit{CD} to 
relatively slower XPUs (e.g., DLAs) reduces GPU contention and improves overall latency.

\section{Fine-grained, Multi-XPU Abstraction}
\label{sec:prog}

\begin{table*}[t]

\vspace{2mm}
\begin{minipage}{1\linewidth}
\small{
\caption{\small{\emph{The main APIs of {\xnode} and {\xmod}.}}}
\label{tab:api}
}    
\end{minipage}
\begin{minipage}{1\linewidth}
\ra{1.1}
\centering
\small{
\begin{tabular}{p{4.5cm} p{8.5cm}}
\toprule
\textbf{API} & \textbf{Description} \\
\midrule
XNode.add\_impl(XPU, init, exec) & Add an implementation to {\xnode}, with XPU type, and callbacks for initialization and execution. \\
XModule.add\_node(XNode) & Add an {\xnode} to {\xmod}. {\xnodes} execute in the order added. \\
XModule.add\_input\_topic(topic) & Add an input topic to {\xmod}, triggering the first {\xnode}. \\
XModule.add\_output\_topic(topic) & Add an output topic to {\xmod}, published by the last {\xnode}. \\
\bottomrule
\end{tabular}
}
\end{minipage} \\[-10pt]
\end{table*}

We present {\xnode}, a fine-grained multi-XPU abstraction that enables 
stage-level XPU management of AuT applications on heterogeneous platforms.
Complementing {\xnode}, we present {\xmod}, a container for 
multiple {\xnodes} that together form a complete algorithm module.
Table~\ref{tab:api} outlines the core APIs of {\xnode} and {\xmod}.
We provide an overview of the abstraction with the help of a simplified code
(see Figure~\ref{fig:code}) that implements camera obstacle detection using {\xnode}
and {\xmod} on NVIDIA Jetson AGX Orin.

\subsection{\xnode}

An {\xnode} serves as the fundamental unit of computation.
It does not represent an entire algorithm, but rather a specific stage 
within an algorithm.
Each {\xnode} can be declared with multiple alternative implementations, 
with each implementation targeting a specific type of XPU.
The key principle in dividing an algorithm into {\xnodes} is that the boundaries
between {\xnodes} should align with transitions between different XPU usages.
For example, if an algorithm uses CPU then GPU, it should be split into two 
separate {\xnodes}, each utilizing a single XPU type.
Through this design, {\xnode} offers two key advantages for AuT applications:

\etitle{Holistic XPU implementation selection.}
An {\xnode} can have multiple XPU implementations, but only one implementation is chosen
for execution, decided by the runtime system.
Programming with {\xnode} allows the developers to focus on the implementation
for each XPU, without worrying about the resource coordination with other {\xnodes}.
The multi-implementation design also provides opportunities for the runtime system to 
make holistic decisions on which implementation to use for each {\xnode} automatically, with the
global view of all {\xnodes} in the application.

\etitle{XPU-aware fine-grained scheduling.}
Each XPU implementation of an {\xnode} is associated with a specific XPU type,
so the runtime system can be aware of the XPU usage of an {\xnode} 
and make scheduling decisions accordingly.
By operating at stage boundaries rather than module boundaries,
this fine-grained approach enables more precise control over XPUs.
Furthermore, since each {\xnode} declares its XPU requirements explicitly,
the runtime system can schedule different XPUs independently while maintaining
a global view of resource utilization across the entire application.

\begin{figure}

    \begin{minipage}{1\linewidth}
        \centering\includegraphics[width=\linewidth]{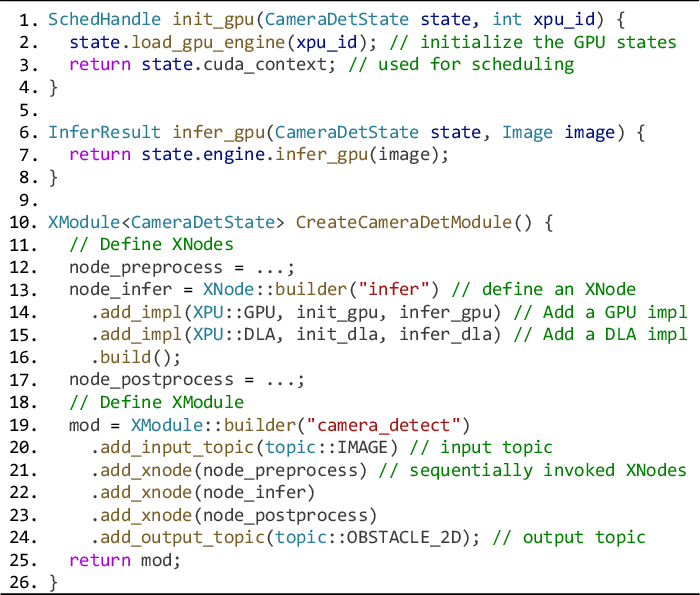} 
    \end{minipage} \\[8pt]
    \begin{minipage}{1\linewidth}
        \caption{\emph{\small{Simplified code that implements a camera
         obstacle detection module using the abstraction of {\xnode} and {\xmod}.}}}
    \label{fig:code}
    \end{minipage} \\[-10pt]
\end{figure}

\stitle{API}.
The primary interface of {\xnode} is the \code{add\_impl} API, which enables the addition of 
alternative implementations to an {\xnode}.
Each implementation requires two key components: a specified XPU type and a pair of callback 
functions for initialization and execution phases.
The initialization callback executes once during {\xnode} setup, performing necessary 
preparation tasks on the specified XPU and returning a \textit{scheduling handle}.
The execution callback implements the core computation logic and is invoked whenever 
the {\xnode} is scheduled for execution on its designated XPU.
The combination of XPU type and scheduling handle is crucial for XPU-aware 
scheduling. The explicit XPU type declaration enables the runtime system to track 
resource utilization, while the scheduling handle provides fine-grained control over 
{\xnode} execution. Together, they enable the runtime system to maintain precise 
control over XPU resources, enabling sophisticated scheduling strategies.

\stitle{Example}.
As shown in Figure~\ref{fig:code} (Lines~1{--}17), the algorithm is 
splited into three {\xnodes}, \code{node\_preprocess}, \code{node\_infer} and \code{node\_postprocess},
each representing a stage of the algorithm.
The \code{infer} XNode is defined with two implementations: one for GPU and one for DLA.
The GPU implementation has an initialization function (\code{init\_gpu}) that 
establishes the GPU context using the specified \code{xpu\_id}, loads the data, and returns 
a CUDA context as the scheduling handle.
The execution function (\code{infer\_gpu}) performs the actual model inference on the GPU.

\subsection{{\xmod}}

An {\xmod} functions as a container for multiple {\xnodes} that collectively form a 
complete algorithm module, analogous to a ROS node or a CyberRT component.
The {\xnodes} within an {\xmod} execute sequentially, following the algorithm's logical flow.
To facilitate efficient data sharing and coordination between algorithm stages, 
the {\xmod} maintains a shared state accessible to all its {\xnodes}.

Similar to other publish/subscribe systems like ROS, {\xmod} implements a 
topic-based communication model to enable inter-{\xmod} communication and loose coupling.
An {\xmod} can subscribe to multiple input topics, with computation initiation 
contingent upon the availability of fresh data across all subscribed topics.
The execution chain begins with the first {\xnode}, which processes the input data 
and passes its output to subsequent {\xnodes} in sequence. The final {\xnode} in 
the chain publishes its results to designated output topics. This topic-based 
architecture naturally expresses dependencies between {\xmods}, where an {\xmod} 
becomes dependent on another when it subscribes to the latter's output topics.

\stitle{Example}. 
Figure~\ref{fig:code} shows how to construct an {\xmod} for the camera
obstacle detection module (Lines~19{--}25). It specifies the input topic, the sequence of {\xnodes}
and the output topic.
The {\xmod} has a shared state \code{CameraDetState}, which maintains the model 
parameters and the intermediate results of the algorithm.

In summary, {\xnode} provides fine-grained control over XPUs 
within algorithm modules, while {\xmod} manages the higher-level dependencies
between modules. Together, these abstractions provide the runtime system with
comprehensive information for XPU coordination and scheduling.

\section{{\sys} Overview}
\label{sec:sys}
We present {\sys}, a framework for deploying and scheduling AuT applications
on heterogeneous hardware platforms that implements the {\xnode} abstraction.
{\sys} efficiently coordinates multiple XPUs to optimize 
end-to-end performance through two key mechanisms.

\stitle{Holistic XPU assignment.}
{{\xnodes} support multiple XPU implementations, with {\sys} performing 
automatic XPU selection to optimize end-to-end performance.
The selection is made holistically across all {\xnodes},
incorporating execution profiles, inter-node dependencies, and XPU resources.
}

\stitle{Fine-grained XPU scheduling}.
{{\sys} implements priority-based preemptive scheduling for {\xnodes} 
on their selected XPUs. An {\xnode} is the basic unit of scheduling, and 
each {\xnode} is assigned a priority level, which is determined offline
in conjunction with XPU assignment decisions.}

\begin{figure}[t]

    \begin{minipage}{1\linewidth}
        
		\centering\includegraphics[width=1.\linewidth]{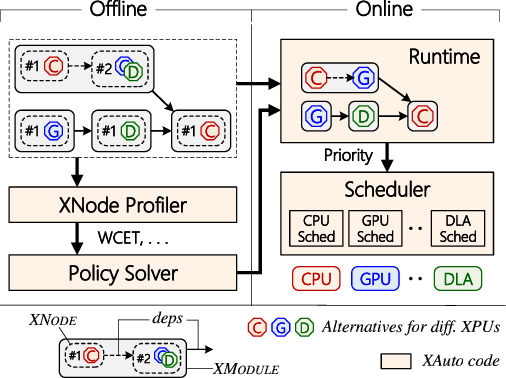}
    \end{minipage} \\[10pt]
    \begin{minipage}{1\linewidth}
        \caption{\emph{\small{The architecture of {\sys}.}}}
    \label{fig:arch}
    \end{minipage} \\[-10pt]
\end{figure}

{\sys} implements these mechanisms through four key components: a profiler, 
policy solver, runtime, and scheduler. The profiler and policy solver 
operate offline during the development phase, while the runtime and 
scheduler function online during execution. Figure~\ref{fig:arch} illustrates 
the workflow.

\stitle{Profiler}.
The development process begins with developers implementing their algorithms using
{\xnode} and {\xmod} abstractions. {\sys}'s profiler then assesses algorithm
performance by executing it against a pre-recorded input trace in an offline
environment. Utilizing the multi-XPU feature of {\xnode}, the profiler accurately
measures individual {\xnode} execution times across different XPUs, including their
worst-case execution times (WCET). To ensure accuracy, the profiler disables concurrent 
{\xnode} execution on each XPU, eliminating any interference effects. This profiling 
process results in comprehensive performance profiles for each {\xnode} across all 
available XPUs, which are essential for making reliable scheduling decisions.

\stitle{Policy solver}.
The policy solver determines XPU assignments and 
priority configurations for each {\xnode} through an Integer Linear 
Programming (ILP) model, as detailed in $\S$\ref{sec:policy}. 
It utilizes execution profiles from the profiler, {\xnode} topology,
and system resource constraints to generate a comprehensive deployment
configuration that specifies both XPU assignments and priorities for all 
{\xnodes}. The policy solver only needs to be run once for one deployment 
configuration.

\stitle{Runtime}.
The runtime orchestrates the execution of {\xnodes} across XPUs according 
to the deployment configuration generated by the policy solver. 
After initializing {\xnodes} and {\xmods} on their assigned XPUs, it manages 
the execution lifecycle of each {\xnode}. Upon receiving new data on an 
{\xnode}'s input topic, the runtime instantiates a task with a statically-assigned 
priority derived from the deployment configuration and submits it to the scheduler. 
When a task completes, the runtime propagates its output to either subsequent {\xnodes} in 
the processing pipeline or to the designated output topic. 
{\sys} allows {\xmods} to execute either within isolated processes or 
within a shared process context.

\stitle{Scheduler}.
The scheduler manages the execution of {\xnodes} across XPUs using
a partitioned scheduling approach, where each XPU maintains its
own independent scheduler. These XPU-specific schedulers expose a priority-based 
interface that enables the runtime to submit tasks. Priority enforcement varies 
by XPU architecture. Preemptive XPUs can interrupt current tasks to execute
higher-priority ones. Non-preemptive XPUs, on the other hand, utilize priorities to determine
task execution order within their queues.

\section{Fine-grained XPU Scheduling}
\label{sec:sched}

\begin{figure}[t]

    \begin{minipage}{1\linewidth}
        
		\centering\includegraphics[width=1.\linewidth]{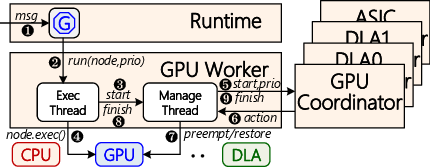}
    \end{minipage} \\[5pt]
    \begin{minipage}{1\linewidth}
        \caption{\emph{\small{The workflow of {\sys}'s XPU scheduler.}}}
    \label{fig:sched}
    \end{minipage} \\[-5pt]
\end{figure}

{\sys} employs a partitioned scheduling approach for XPUs,
where each XPU has a dedicated coordinator to manage the tasks.
Figure~\ref{fig:sched} illustrates the architecture of {\sys}'s scheduler,
and the workflow of task submission and scheduling.

\subsection{Scheduler Architecture}

There are two primary components in the scheduling workflow: XPU workers 
and XPU coordinators.

\stitle{XPU workers}. Each {\xmod} maintains dedicated workers for its 
associated XPUs. These workers expose a priority-based task submission interface
and operate using two distinct threads: an execution thread and a management thread.
The execution thread handles the invocation of the {\xnode}'s \textit{exec} function,
while the management thread performs task preemption and restoration based on
the coordinator's scheduling decisions. The implementation of preemption
and restoration are XPU-dependent (detailed in $\S$~\ref{sec:impl}).

\stitle{XPU coordinators}. Each XPU instance has a centralized coordinator,
implementing a fixed-priority scheduling policy.
The coordinator runs in a separate process, and interacts with the 
XPU workers through inter-process communication (IPC).
It receives the events of task start and finish from the 
XPU workers, and decides which scheduling action to take based on 
the task priority and the current scheduling status of the XPU.
The implementation of the coordinator is XPU-agnostic, interacting
solely with the XPU workers without directly touching the XPU itself.

\stitle{Scheduling Workflow.}
When an {\xnode} receives a new message (\ding{202}), the runtime system 
submits the corresponding task to the appropriate XPU worker with its 
associated priority (\ding{203}). Upon receiving the task, the execution 
thread signals the task initiation to the management thread (\ding{204}) 
before invoking the {\xnode}'s \textit{exec} function (\ding{205}). 
The management thread then notifies the XPU's coordinator of the 
task status (\ding{206}). Based on the task's priority and the XPU's 
current scheduling state, the coordinator determines the appropriate 
scheduling action. If a task with higher priority is already executing, 
the coordinator instructs the worker to suspend the current task (\ding{207}), 
prompting the management thread to invoke the XPU-specific \textit{preempt} 
function (\ding{208}). Conversely, if the current task has the highest 
priority, the coordinator directs the worker to resume execution through 
the XPU-specific \textit{restore} function. Upon task completion, the 
execution thread notifies the management thread (\ding{209}), which then 
relays this information to the coordinator (\ding{210}).

\subsection{XPU Worker Implementations}
\label{sec:impl}

The XPU worker needs to implement the preempt and restore mechanisms
to achieve preemptive scheduling on the XPU. 
The mechanisms are XPU-specific, taking different forms on different XPUs. 

\stitle{GPU worker}. 
NVIDIA GPUs employ a default round-robin time-sharing scheduling policy,
which utilizes Time Slice Groups (TSGs) to distribute computational resources
among processes~\cite{Wang2024UnleashingTP}. Each CUDA Context is typically
mapped to a distinct TSG, and the NVIDIA GPU driver exposes interfaces to enable
and disable these TSGs. When a TSG is disabled, its associated GPU tasks are
preempted; conversely, enabling a TSG allows tasks to resume execution.
Based on this mechanism, we implement the \textit{preempt} and \textit{restore}
functions by invoking the TSG disable and enable operations, respectively.
To utilize this functionality, applications must provide their CUDA context
(\code{CUcontext}) as a scheduling handle during {\xnode} initialization
(see Figure~\ref{fig:code}). This allows the GPU worker to identify the
corresponding TSG and manage task preemption and restoration.

\stitle{DLA worker}.
NVIDIA DLA also implements a default round-robin 
time-sharing scheduling policy for task management. Within the kernel-space 
driver, each process maintains its own command queue. The kernel driver 
provides interfaces to suspend and resume these command queues, which {\sys} 
utilizes to implement the \textit{preempt} and \textit{restore} functions, 
respectively. To enable this scheduling mechanism, applications must provide 
their \code{cudlaDevHandle} as a scheduling handle during {\xnode} 
initialization, allowing the system to identify and manage the corresponding 
command queue.

\stitle{Default XPU workers}.
For XPUs lacking native preemption capabilities, {\sys} implements a 
non-preemptive priority-based scheduling mechanism. In this approach, 
task execution follows a strict priority order enforced through 
coordination between workers and the scheduler. Upon task submission, 
the worker signals the coordinator and suspends the submission process 
until receiving an execution approval. The coordinator maintains the 
global scheduling state and determines task execution order based on 
their assigned priorities, ensuring higher-priority tasks are processed 
first.

\stitle{CPU worker}.
Unlike other XPU implementations, the CPU worker leverages the native 
priority-based preemptive scheduling capabilities provided by the Linux 
kernel~\cite{LinuxSched}. For each CPU worker, {\sys} creates a dedicated 
Linux thread and assigns it a priority level matching the task priority. 
The thread operates under the SCHED\_FIFO scheduling policy and is bound 
to a CPU core.

\section{ILP-based Scheduling Policy}\label{sec:policy}

{\sys} determines the XPU assignment and the priority of each {\xnode}
before the runtime starts.
The XPU assignment determines which XPU should be used to execute 
each {\xnode} (not only the XPU type, but also the specific XPU ID),
while the priority settings determine the execution order of different {\xnodes}
on the selected XPUs.
However, obtaining an efficient scheduling configuration on these two 
aspects is non-trivial.
It is challenging as many factors need to be considered, including 
the concurrency and dependency of different {\xnodes}, the performance and 
contention of different XPUs, etc.
These factors are interrelated and may lead to contradicting decisions.
For example, assigning an {\xnode} to use GPU may achieve higher performance
than using NPU, but assigning too many {\xnodes} to GPU may lead to contention
and finally degrade the overall performance.

\stitle{Limitations of brute-force and heuristic approaches.}
A straightforward approach involves performing a brute-force search of all
possible XPU assignments and priority settings, subsequently selecting the
configuration that minimizes end-to-end latency in a scheduling simulation.
However, the search space expands exponentially with the number of {\xnodes} and
XPUs, leading to significantly prolonged search times.
For instance, with $N$ {\xnodes} and $M$ XPUs, there are $M^N$ possible XPU
assignments and $N!$ possible priority settings.
Our experiments show that even with $N=20$ and $M=6$, the search time can
reach more than 24 hours.
An alternative method is to employ a heuristic strategy, such as 
HEFT~\cite{Topcuoglu2002PerformanceEffectiveAL}. Nevertheless, our experiments
reveal that heuristic approach can result in sub-optimal scheduling performance,
thereby limiting the full potential of XPUs.

\stitle{Our approach.}
To obtain an efficient scheduling configuration within a acceptable 
time (e.g., minutes), we propose to construct an Integer Linear 
Programming (ILP) model to find the XPU assignment and priority setting 
for each {\xnode}.
The ILP model can combine all the factors to find a near-optimal solution,
mitigating the scheduling performance issue of the heuristic approach.
Moreover, the ILP model can be efficiently solved using off-the-shelf solvers,
directly benefiting from various acceleration techniques of these solvers (such as
pruning, parallelization, etc.).

\begin{table}[t]
    \vspace{3mm}
    \begin{minipage}{1.\linewidth}
    \caption{\small{\emph{Variables used in the ILP model.}}}
    \label{tab:var}
    \end{minipage} 
    \begin{minipage}{1.\linewidth}
    \ra{1.1}
    \centering
    \small{
    
    \begin{tabular}{@{~}l@{~~~~~~}l@{~~}l@{~~~~~~}l@{~}}
    \toprule
    & \textbf{Variable} & \textbf{Type} & \textbf{Description} \\
    \midrule
    & $s_{i}$ & Real & Start time of {\xnode} $v_{i}$ \\
    & $f_{i}$ & Real & Finish time of {\xnode} $v_{i}$ \\
    & $d^m_{i,j,k}$ & Real & Time demand of $v_{k}$ on $m$ in interval $[s_{i}, f_{j}]$ \\
    & $p_{i}^{m}$ & Binary & {\xnode} $v_i$ is assigned to XPU $m$ if $p_{i}^{m}=1$ \\
    \bottomrule
    \end{tabular}
    }
    \end{minipage} \\[-8pt]
\end{table}

\stitle{Key ideas.}
Our ILP model constructs a time table to schedule all {\xnodes}, where each {\xnode}'s start 
and finish times are model variables. The objective is to minimize the maximum finish 
time (end-to-end latency). The schedule must satisfy three constraints: (1) precedence 
constraints ensure correct execution order; (2) WCET constraints guarantee sufficient execution
time on assigned XPUs; and (3) resource constraints prevent XPU contention. 
The ILP solver combines these constraints with the objective function 
to find a near-optimal schedule.

\stitle{Problem statement.}
Table~\ref{tab:var} shows the variables used in the model.
Given a directed acyclic graph $G = (V, E)$ and a set of XPUs $M$,
where $V$ is the set of {\xnodes} and $E$ is the set of directed edges
representing precedence between {\xnodes}. 
Each {\xnode} $v_i \in V$ has WCETs $\{wcet_{i}^{m}\}_{m=1}^M$ on each XPU $m \in M$.
If an {\xnode} $v_i$ cannot be executed on XPU $m$, $wcet_{i}^{m}=\infty$.
All {\xnodes} in $G$ should complete within deadline $D$.
The goal is to assign each {\xnode} $v_i \in V$ to an XPU $m \in M$ and 
determine the priority of each {\xnode} to ensure each {\xnode} $v_i$ completes
within the deadline.

The constraints and objective function of our ILP model are as follows:

\etitle{Processor assignment.}
We first introduce a set of binary variables $p_{i}^{m}$ to indicate whether
{\xnode} $v_i$ is assigned to XPU $m$. If an {\xnode} $v_i$ does not support
XPU $m$, $p_{i}^{m}$ is constantly set to 0.
To ensure that each {\xnode} is assigned to exactly one XPU, for each ${v_i}$
we introduce the following constraint:
\begin{equation}
    \sum_{m} p_{i}^{m} = 1 
\end{equation}

We next introduce the non-negative real-valued variables $s_{i}$ and $f_{i}$ to represent
the start and finish time of {\xnode} $v_i$.
To ensure the validity of these times, we have the following constraints:

\etitle{Enforcing precedence.}
Because {\xnodes} have dependencies with each other, we need to ensure that the start time
of $v_i$ is later than the finish time of its predecessors (i.e., the
{\xnodes} that have edges to $v_i$).
For each {\xnode} $v_i$ and each of its predecessors $v_j$, we introduce
the following constraint:
\begin{equation}
   f_{j} \leq s_{i} 
\end{equation}

\etitle{Enforcing WCET.}
Due to the support for task preemption by XPUs, an {\xnode} may not be executed
continuously on a XPU.
Consequently, the interval between the start and finish times of each {\xnode} must
be at least as long as its WCET.
Since an {\xnode} can have different WCETs on different XPUs, we need to select the
WCET corresponding to the XPU to which the {\xnode} is assigned, 
denoted by $p_{i}^{m}$. 
Specifically, for each {\xnode} $v_i$ and each XPU $m$, we introduce the following constraint:
\begin{equation}
    f_{i} \geq s_{i} + wcet_{i}^{m} \quad\text{if}\quad p_{i}^{m} = 1
\end{equation}

The above constraint is not linear as it contains a conditional
expression, but it can be linearized by introducing 
auxiliary constraints. For simplicity, we omit the linearization representation.

\etitle{Enforcing schedule feasibility.}
We next ensure that each XPU can only execute one {\xnode} 
at a time. Inspired by prior work~\cite{Baruah2020SchedulingDW}, we also leverage the 
demand-bound function characterization~\cite{PreemptivescBaruah90} to model such constraint. 
It has been proven that a sufficient and necessary condition for a feasible 
schedule of a set of tasks on the same XPU is that for any interval $[s, f]$ where $s$ is 
the start time of some task and $f$ is the finish time of some task, the total required 
processing time of all tasks on the XPU within the interval is no more than $f - s$.
We model this constraint with two steps. 
First, we introduce the real-valued variables $d_{i,j,k}^{m}$,
which is intended to represent the amount of processing time requested by {\xnode} $v_k$
on XPU $m$ over the interval $[s_{i}, f_{j}]$.
To enforce this interpretation, for each triple of {\xnodes} $v_i$, $v_j$ 
and $v_k$ and each XPU $m$, we introduce the following constraints:

\begin{equation}
    \begin{aligned}
        d_{i,j,k}^{m} \geq wcet_i^m ~~~~~\text{if}~~~ &(p_{i}^{m} = p_{j}^{m} = p_{k}^{m} = 1)\\
                                                     &\wedge (s_{i} \le s_{k}) \wedge (f_{j} \ge f_{k})
    \end{aligned}
\end{equation}

Second, we need to ensure that for each valid interval
$[s_{i}, f_{j}]$ (i.e., $s_{i} \le f_{j}$), the accumulated demand of all nodes on the XPU $m$ does not exceed
its capacity (i.e., $f_{j} - s_{i}$):
\begin{equation}
    \sum_{k}{d_{i,j,k}^{m}} \leq f_{j} - s_{i} \quad\text{if}\quad (s_{i} \le f_{j})
\end{equation}

The conditions for the above constraints can also be linearized using 
auxiliary binary variables.

\etitle{Minimizing end-to-end latency.}
With the above constraints, we can build a valid schedule where 
each {\xnode} is assigned to an XPU and the schedule is feasible under 
preemptive scheduling.
Finally, we can use the following objective function to minimize 
the end-to-end latency:
\begin{equation}
    \min \max_{i=1}^N f_{i}
\end{equation}

\stitle{Interpretation of the variables.}
After solving the ILP model, we can then map the model variables to the
XPU assignment and priority settings.
The XPU assignment is straightforward as the $p_{i}^{m}$ directly indicates which 
XPU is assigned to each {\xnode}.
As for the priority settings, we can use the $f_{i}$ to determine the 
priority of each {\xnode}. Specifically, we sort the {\xnodes} by their 
finish time $f_{i}$ and assign higher priority to the {\xnode} with 
earlier finish time.

\section{Evaluation}\label{sec:eval}

We conducted our evaluation on an NVIDIA Jetson AGX Orin development platform,
focusing on the scheduling of CPU, GPU and DLA as they are the most common 
XPUs in AuT applications.
To ensure consistent performance measurements, we configured the
platform to operate in MAXN power mode with all processors running at their 
maximum clock frequencies.

\subsection{Overall Performance of {\sys}}
\label{sec:eval-overall}

We first evaluate the performance of {\sys} using synthetic 
workloads to assess its capabilities under various conditions.

\stitle{Synthetic workloads.}
Our evaluation employs a set of synthetically generated DAGs created 
through a two-phase process. In the first phase, we construct DAGs containing 
specified numbers of {\xnodes} using established DAG generation 
algorithms~\cite{Nasri2019ResponseTimeAO,Melani2015ResponseTimeAO}. The second 
phase augments each {\xnode} with implementations for randomly selected XPU types. 
The evaluation dataset consists of 7 groups, each containing 50 unique DAGs, with 
the average number of {\xnodes} per group ranging from 10 to 40 in increments of 5.

\etitle{DAG construction.}
The DAG structure is constructed using a recursive series-parallel generation approach.
Starting with a single node, the algorithm recursively expands each node into either
a terminal node or a parallel sub-graph with a predefined probability (0.4 for a 
terminal node, 0.1 for adding random edges between sibling nodes).

\etitle{{\xnode} construction.}
Each DAG node represents an {\xnode} with a baseline CPU execution time uniformly 
sampled from 5--95\,ms (50\,ms on average).
Implementations for GPU and DLA are then generated following the principle that
the tasks with longer CPU execution times are more likely to 
have GPU and DLA implementations.
Specifically, we categorize {\xnode}s into a group with the top 60\%
longest CPU execution times and another group with the bottom 40\%.
GPU implementations are added with probability 1.0 to the top group 
and 0.5 to the bottom group (0.8 on average), with 
speedups of 3$\times$--8$\times$ relative to the CPU execution time. 
DLA implementations follow the same pattern: 
0.8 probability for the top group and 0.3 for the bottom group (0.6 on average), 
with speedups of 2$\times$--5$\times$.

\stitle{Experiment setup.}
With the generated DAGs and {\xnode} implementations, we evaluate 
the performance of {\sys} in the following approaches.

\etitle{XPU configurations.}
For each test case, we evaluate its performance on two XPU configurations,
simulating different resource constraints.
A configuration (\textbf{small config}) uses 4 CPUs (cores), 1 GPU and 1 DLA,
and another (\textbf{large config}) uses 8 CPUs (cores), 1 GPU, 2 DLAs.

\etitle{XPU task execution.}
Task execution is simulated by repeating a basic execution unit for each XPU type.
For CPU implementations, a floating point multiplication serves as the basic unit, while GPU 
implementations use a spinning kernel, and DLA implementations employ 
a convolution operator. The number of repetitions for each basic unit is profiled 
to match the assigned execution time of the corresponding {\xnode} implementation.

\etitle{Comparing targets.}
{We compare {\sys} against \textbf{ROS2}~\cite{ROS2Paper}, a state-of-the-art runtime system 
widely adopted in AuT applications. Since ROS2 lacks native support for {\xnode} 
execution and XPU implementation selection, we manually pre-assign XPUs to each {\xnode} 
for ROS2 using a local-optimal heuristic, i.e., selecting the fastest available XPU for each {\xnode} 
independently. This policy is reasonable for ROS2 as it does not provide a global view of the XPU usage, 
thus can only perform local optimization.
For a more comprehensive comparison, we introduce \textbf{{\sysH}}
which implements the Heterogeneous Earliest Finish Time (HEFT) 
algorithm~\cite{Topcuoglu2002PerformanceEffectiveAL} on top of {\sys}.
HEFT is a widely-used heuristic approach for assigning tasks to heterogeneous 
processors for a DAG-based application.

}

\begin{figure}[t]
    \begin{minipage}{\linewidth}
        \centering\includegraphics[width=\linewidth]{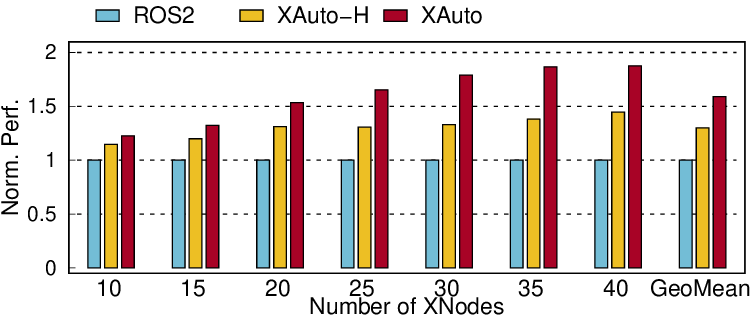}
    \end{minipage} \\[-8pt]
    \begin{minipage}{\linewidth}
        \centering\includegraphics[width=\linewidth]{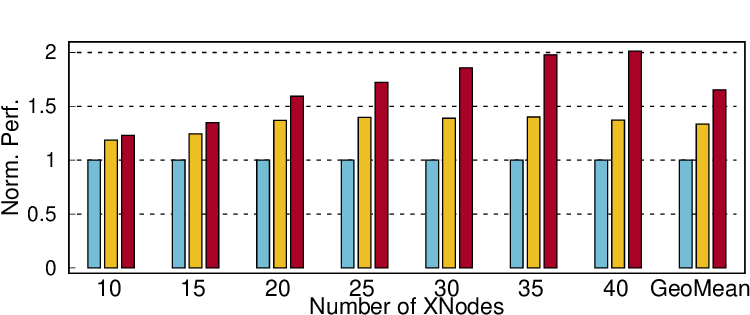}
    \end{minipage} \\[8pt]
    \begin{minipage}{1.\linewidth}
        \caption{\emph{\small{The performance of {\sys} and ROS2 
        under (Top) \textbf{small config} and (Bottom) \textbf{large config}, 
        increased with the number of {\xnodes} in the DAGs.
        Experiments are conducted on an NVIDIA Jetson AGX Orin development platform.}}}
    \label{fig:eval-bench}
    \end{minipage} \\[-5pt]
\end{figure}

\begin{figure}[t]
    \begin{minipage}{\linewidth}
        \centering\includegraphics[width=\linewidth]{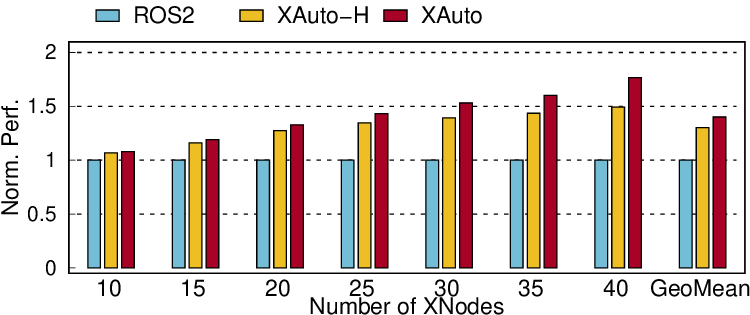}
    \end{minipage} \\[8pt]
    \begin{minipage}{1.\linewidth}
        \caption{\emph{\small{The performance of {\sys} and ROS2 
        under \textbf{small config}, 
        increased with the number of {\xnodes} in the DAGs.
        Experiments are conducted on an Intel Core Ultra 7 155H platform.}}}
    \label{fig:eval-intel}
    \end{minipage} \\[5pt]
\end{figure}

\begin{figure}[t]
    \begin{minipage}{\linewidth}
        \centering\includegraphics[width=\linewidth]{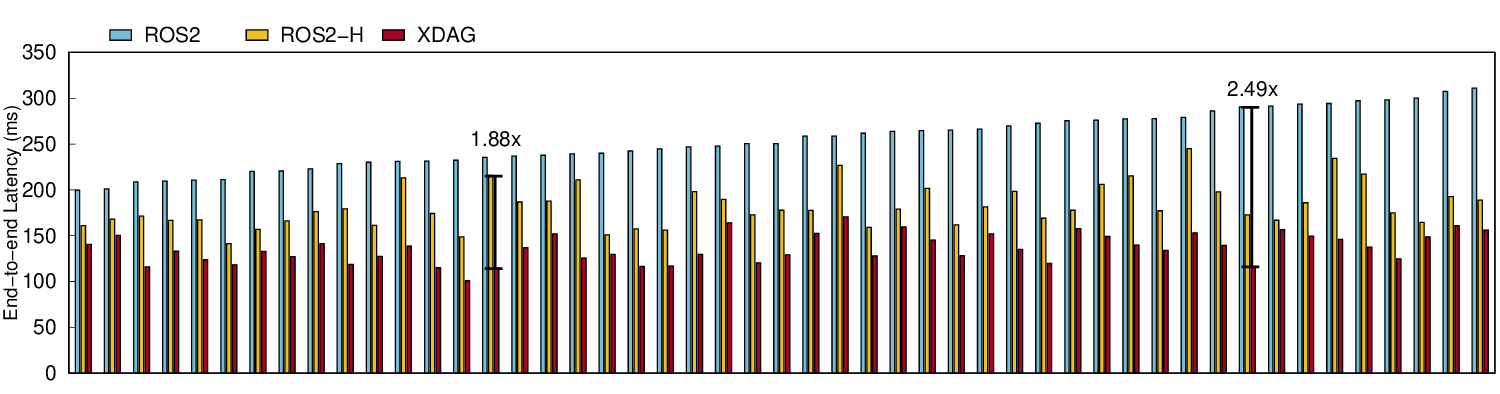}
    \end{minipage} \\[5pt]
    \begin{minipage}{1.\linewidth}
        \caption{\emph{\small{The performance of {\sys} and ROS2 
        using a group of DAGs with 30 {\xnodes}. 
        Experiments are conducted on an NVIDIA Jetson AGX Orin development platform.}}}
    \label{fig:eval-bench}
    \end{minipage} 
\end{figure}

\stitle{Results.}
Figure~\ref{fig:eval-bench} illustrates the performance comparison between {\sys} and ROS2 
across two XPU configurations, with DAG sizes ranging from 10 to 40 {\xnodes}. 
Each DAG's performance is the average end-to-end latency normalized to ROS2's result, 
and each group's performance is the geometric mean of all DAGs.

\etitle{{\sys} consistently outperforms ROS2.}

Both {\sys} and {\sysH} achieves notable performance improvements over ROS2.
Specifically, {\sys} significantly and consistently outperforms ROS2,
achieving 1.23$\times$--1.88$\times$ (1.59$\times$ on GeoMean) improvements for small config
and 1.23$\times$--2.01$\times$ (1.65$\times$ on GeoMean) improvements for large config.
Notably, the foundation of the performance improvements for both {\sys} and {\sysH} lies in the 
global view of XPU utilization enabled by the {\xnode} abstraction,
highlighting the importance of such a multi-XPU abstraction.

\etitle{{\sys} shows more performance gain than {\sysH}.}
Although both {\sys} and {\sysH} assign XPUs holistically, 
{\sys} shows more performance gain than {\sysH}.
Specifically, {\sys} outperforms {\sysH} by 1.22$\times$ and 1.24$\times$
for small and large config, respectively.
This improvement is primarily attributed to {\sys}'s ILP-based XPU assignment, 
which explores more possible XPU assignments than the heuristic approach in {\sysH}.

\etitle{{\sys} shows more advantages with larger DAGs.}
The performance advantage of 
{\sys} becomes increasingly pronounced as DAG size increases compared to ROS2. While 
{\sys} achieves a modest 1.23$\times$ improvement over ROS2 with 10 {\xnodes}, this 
advantage grows substantially to 2.01$\times$ with 40 {\xnodes} under the large config.
This performance differential stems from ROS2's tendency to assign tasks to GPU,
leading to increased resource contention in larger DAGs. In contrast, {\sys} effectively leverages its ILP-based
XPU assignment to optimize resource utilization across the system.

\etitle{{\sys} effectively utilizes additional XPU resources.}
The performance improvement of {\sys} grows from 1.88$\times$ with limited resources (small config) 
to 2.01$\times$ when given additional
processors (large config with 4 more CPUs and 1 more DLA) for DAGs with 40 {\xnodes}
compared to ROS2.
While ROS2 struggles to distribute workloads across the processors, 
{\sys} leverages the additional XPUs to further reduce execution time, 
highlighting the effectiveness of {\sys}'s holistic XPU assignment.

\subsection{Effectiveness of XPU Assignment and Scheduling}

\begin{figure}
    \begin{minipage}{\linewidth}
        \centering\includegraphics[width=\linewidth]{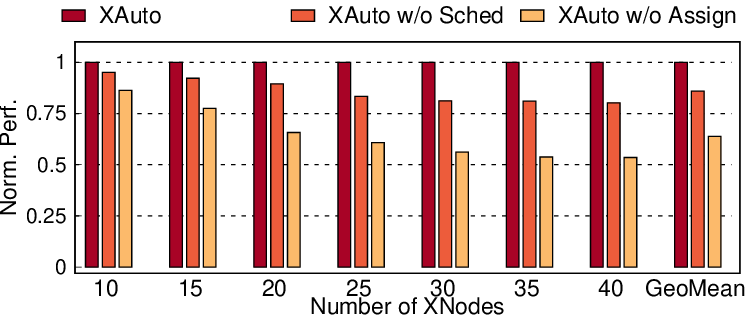}
    \end{minipage} \\[-8pt]
    \begin{minipage}{\linewidth}
        \centering\includegraphics[width=\linewidth]{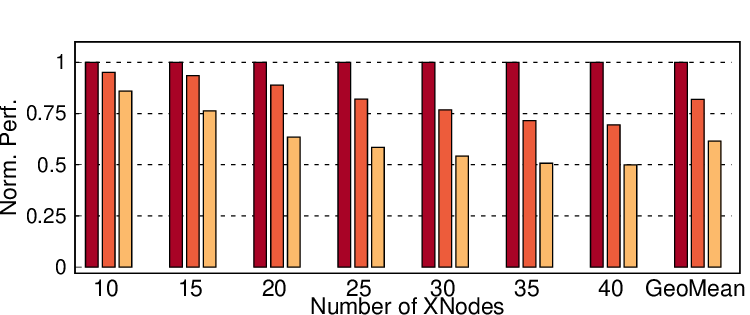}
    \end{minipage} \\[8pt]
    \begin{minipage}{1.\linewidth}
        \caption{\emph{\small{The performance of {\sys} without XPU scheduling or XPU assignment
        under (Top) \textbf{small config} and (Bottom) \textbf{large config}, increased with the number of 
        {\xnodes} in the DAGs.}}}
    \label{fig:eval-ablation}
    \end{minipage} \\[-20pt]
\end{figure}

We next analyze the effectiveness of the two key mechanisms 
of {\sys}: XPU assignment and XPU scheduling.

\stitle{Ablation analysis.}
To demonstrate the individual contributions of each mechanism to {\sys}'s overall
performance, we conducted an ablation study using the same synthetic DAGs and 
XPU configurations as described in $\S$\ref{sec:eval-overall}.
We examined two variants: \textbf{{\sys} w/o Sched}, where the XPU scheduler is disabled
and scheduling decisions are delegated to the hardware scheduler, and
\textbf{{\sys} w/o Assign}, where the ILP-based XPU assignment is replaced with
the local-optimal heuristic approach which chooses the fastest available XPU for 
each {\xnode}.
Performance measurements for each test case are normalized relative to the baseline
{\sys} implementation with both scheduling and XPU assignment enabled.
Results are presented in Figure~\ref{fig:eval-ablation}.

\etitle{Both XPU scheduling and XPU assignment are essential.}
The ablation results demonstrate that disabling either mechanism leads to significant
performance degradation across all DAG sizes.
Specifically, removing XPU scheduling ({\sys} w/o Sched) results in a 16.1\% slowdown 
on average, while disabling XPU assignment ({\sys} w/o Assign) causes a 37.3\% performance 
drop. This indicates that the two mechanisms are complementary - XPU scheduling optimizes 
resource utilization at runtime, while XPU assignment provides the foundation for efficient 
task distribution.

\etitle{Performance impact grows with DAG size.}
The performance degradation from disabling either mechanism becomes 
more pronounced as DAG size increases. 
With 10 {\xnodes}, disabling scheduling or assignment results in modest 
slowdowns by 4.9\% and 13.9\% respectively. However, 
these penalties increase dramatically with 40 {\xnodes}, where 
{\sys} w/o Sched experiences a 25.4\% slowdown and {\sys} w/o Assign 
shows a 48.3\% performance drop. This escalating impact reflects the 
growing complexity of resource management in larger DAGs, where both holistic 
XPU assignment and fine-grained scheduling become increasingly critical for 
maintaining performance.

\stitle{Performance of XPU schedulers.}
We next analyze the performance of {\sys}'s XPU schedulers.

\etitle{Preemption latency.} 
The GPU and DLA schedulers support preemptive scheduling utilizing the
interface exposed by the hardware scheduler. 
For GPU scheduler, preempting a running task and switching to another task 
together takes 275\,{\us} on average, while DLA scheduler 
requires 132\,{\us}. 

\etitle{CPU utilization.}
The XPU coordinator implements a priority-based scheduling policy that incurs negligible 
computational overhead. The scheduler is invoked only at task boundaries, specifically 
when an {\xnode} begins or completes execution, making it highly efficient. 
Our measurements show in the workflow with 40 {\xnodes}, 
the combined CPU utilization of all XPU schedulers remains below 0.5\%.

\begin{figure}

    \begin{minipage}{.65\linewidth} 
        \centering\includegraphics[width=0.49\linewidth]{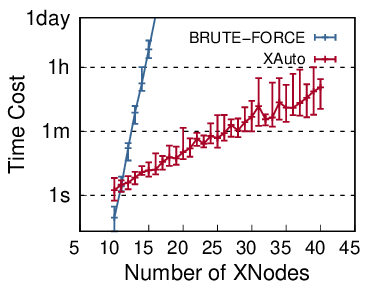} 
        \centering\includegraphics[width=0.49\linewidth]{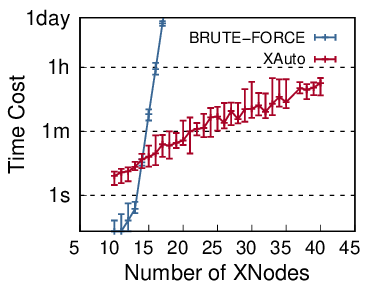} 
        \caption{\emph{\small{The time cost of solving the ILP model and 
        brute-force search all possible assignments under (a) small config and 
        (b) large config, increased with the number of {\xnodes} in DAGs. 
        The error bars show the minimum and maximum solving time.}}}
        \label{fig:eval-solving-time}
    \end{minipage} 
    \hfill
    \begin{minipage}{.33\linewidth} 
        \centering\includegraphics[width=\linewidth]{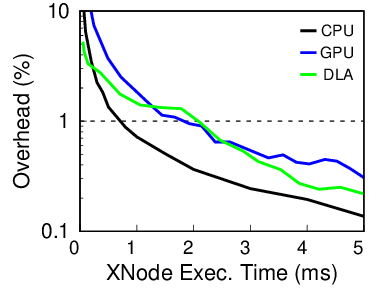} 
        \caption{\emph{\small{Overhead of {\sys}'s runtime, 
        increased with the execution time of {\xnodes}.}}}
    \end{minipage} \\[-5pt] 
\end{figure}

\stitle{Solving time of ILP model.}
We next analyze the computational complexity of solving our ILP model.
Using the Gurobi solver~\cite{gurobi} with 32 parallel threads, we evaluated the model's
solving time across varying numbers of {\xnodes}. Figure~\ref{fig:eval-solving-time} presents
the time cost of solving our ILP model and that of a brute-force search approach.
The results demonstrate that for DAGs containing up to 20 {\xnodes}, our ILP model consistently
converges to the optimal solution within one minute for both small and large configs.
In contrast, the brute-force approach exhibits significantly higher computational demands,
requiring over one hour to evaluate all possible assignments for just 15 {\xnodes}, and
exceeding 24 hours for 20 {\xnodes}. 
Although the solving time of our ILP model still exhibits exponential growth with
respect to the number of {\xnodes}, it has made it practical for typical AuT 
applications with a dozen algorithm modules.

Because the model only needs to be solved once before deployment, such a solving time is 
acceptable for most cases.
For larger-scale applications, approximate methods can be applied to reduce the solving
time, such as increasing the optimal gap for early termination.

\subsection{Case Study}

We further study the performance of {\sys} using an autonomous driving application.
This application, named \textit{Driving Case}, is constructed with reference 
to the perception component of open-source autonomous driving systems, 
including the Autoware.Auto~\cite{AutowareGithub} and the Apollo~\cite{Apollo}.

\begin{figure}[t]
    \begin{minipage}{.30\linewidth}
		
		\hspace{5mm}
        \centering\includegraphics[width=.95\linewidth]{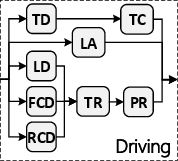} 
    \end{minipage}
	\hspace{0mm}
    \begin{minipage}{.70\linewidth}
		\centering\includegraphics[width=.98\linewidth]{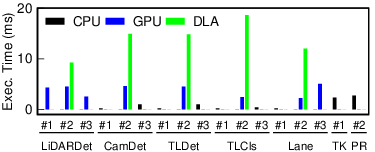} 
    \end{minipage} \\[10pt]    
    \begin{minipage}{1\linewidth}
        \caption{\emph{\small{(a): The topological relationships between {\xmods} of 
        the \textit{Driving Case}. (b): The execution time of each {\xnode} implemented on 
        XPUs. \#n means the n-th {\xnode} in the {\xmod}.}}}
    \label{fig:eval-dag-case}
    \end{minipage} \\[-5pt]
\end{figure}

\stitle{Workload.}
The \textit{Driving Case} processes input from 
three sensors: two cameras (front and rear) and one LiDAR sensor, which feed into 
three parallel processing pipelines. The application topology is illustrated in 
Figure~\ref{fig:eval-dag-case}~(a). The front camera stream undergoes two parallel 
processes: lane detection module (\textit{LA}) using the LaneNet model~\cite{wang2018lanenet},
and traffic light detection module (\textit{TD}) using the YOLOX model~\cite{ge2021yolox}. 
The detected traffic lights are further analyzed by a traffic light classification module 
(\textit{TC}) utilizing the MobilenetV2 model~\cite{sandler2019mobilenetv2}.
Concurrently, obstacle detection is performed through three parallel paths: two camera 
detection modules (\textit{FCD} and \textit{RCD}) using YOLOv3~\cite{redmon2018yolov3}, 
and a LiDAR detection module (\textit{LD}) employing the PointPillars 
model~\cite{lang2019pointpillars}. While the camera detection modules share the same 
processing flow, they operate with distinct model parameters. The detected obstacles 
from all sources are then merged and processed through tracking 
(\textit{TR}) and prediction (\textit{PR}) modules to analyze obstacle trajectories.

\stitle{Implementation on {\sys}.}
The implementation utilizes three XPU types (CPU, GPU, and DLA).
Multi-XPU implementations (GPU and DLA) are specifically employed for model inference {\xnodes},
with TensorRT~\cite{TensorRT} handling the compilation and execution of all models. 
Due to DLA's architectural constraints, including limited support for certain DNN layers and 
restrictions on input and weight dimensions, we employ two mitigation strategies: 
(1) architectural adaptation of DNN models to accommodate DLA requirements (e.g., decomposing 
single MaxPool operations into two sequential MaxPool operations with kernel size 5), and 
(2) model partitioning across multiple {\xnodes}, enabling partial DLA execution where supported.
Each {\xmod} consists of one to three {\xnodes}, with their execution times illustrated in 
Figure~\ref{fig:eval-dag-case}~(b). Notably, the GPU implementation consistently delivers 
superior performance across all {\xnodes}.

\begin{figure*}
    \begin{minipage}{1\linewidth}
        \centering\includegraphics[width=\linewidth]{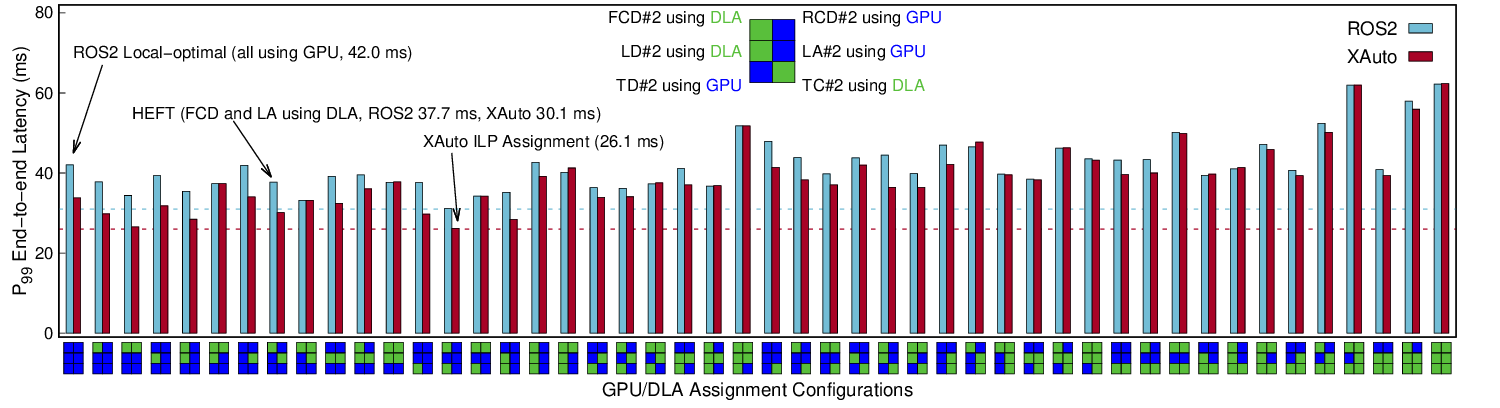} 
    \end{minipage}  \\[5pt]
    \begin{minipage}{1\linewidth}
        \caption{\emph{\small{The end-to-end latency of the Driving Case 
         under different GPU/DLA assignments using ROS2 and {\sys}. 
         The x-axis represents the GPU/DLA assignment, where the blue block 
         and green block means assigning GPU and DLA for the corresponding {\xnode}, 
         respectively. The blue and red dashed lines represent the lowest latency 
         achieved by ROS2 and {\sys}, respectively.}}}
    \label{fig:eval-xdag-sched}
    \end{minipage} \\[-5pt]
\end{figure*}

We deployed the \textit{Driving Case} on both {\sys} and ROS2.
ROS2 uses the same XPU assignment policies outlined in $\S$\ref{sec:eval-overall},
i.e., choosing the fastest XPU for each {\xnode}.
Specifically, in the \textit{Driving Case}, ROS2 uses GPU for all model inference {\xnodes}.
{\sys} uses the ILP model to find the XPU assignment,
which finally uses DLA for the model inference {\xnodes} in the \textit{FCD} and \textit{TD} modules,
with GPU handling the rest. 

The results demonstrated {\sys}'s superior performance: ROS2 achieved 99-percentile end-to-end 
latency of 42.0\,ms, while {\sys} achieved 26.1\,ms, representing an improvement of 1.61$\times$.
This performance gain can be attributed to {\sys}'s combination of precise XPU 
assignment and fine-grained XPU scheduling capabilities.

To further demonstrate the effectiveness of {\sys}'s dual optimization approach combining 
XPU assignment and scheduling, we conducted an exhaustive evaluation comparing {\sys} and 
ROS2 across all possible GPU/DLA assignment configurations, as illustrated in 
Figure~\ref{fig:eval-xdag-sched}. 
The results showed that {\sys} achieved 
lower end-to-end latency compared to ROS2 for most assignments, with improvement 
up to 1.29$\times$. 
This performance enhancement can be attributed to {\sys}'s fine-grained XPU
scheduling mechanism. 
For instance, when examining the configuration where all inference tasks 
execute on GPU (ROS2's local-optimal assignment), {\sys} strategically prioritizes \textit{LD}, \textit{FCD}, and 
\textit{RCD} over \textit{TD}. This prioritization enables parallel execution of 
CPU-dominated tasks (\textit{TR} and \textit{PR}) alongside GPU-dominated tasks 
(\textit{TD} and \textit{TC}), thereby maximizing resource utilization.
Furthermore, {\sys}'s GPU scheduler maintains single-task GPU occupancy, eliminating 
the substantial overhead associated with context switching that occurs in default 
round-robin scheduling.

Notably, our comprehensive evaluation across all possible assignment configurations revealed 
that both ROS2 and {\sys} achieved their optimal end-to-end latency using the assignment 
policy determined by the ILP model (i.e., using DLA for \textit{FCD} and \textit{TD}).
For comparison, when using the HEFT algorithm-based assignment, which allocated DLA to 
model inference {\xnodes} in the \textit{FCD} and \textit{TD} modules, {\sys} achieved 
a 30.1\,ms end-to-end latency, which is worse than the 26.1\,ms achieved with ILP-based 
assignment. These results further validate the effectiveness of our ILP-based approach 
in identifying optimal XPU assignment.

\section{Discussion and Future Work}

\nostitle{Advantages of static scheduling}.
{\sys} uses static scheduling where XPU assignments and priorities are determined before runtime 
based on offline profiling. This approach suits workloads with stable execution patterns, 
such as autonomous driving applications where ML models have consistent latency characteristics. 
The static approach offers key advantages: our ILP-based policy provides deterministic performance bounds 
through WCET analysis, crucial for latency-sensitive applications where deadline misses can cause system failures. 
Static scheduling also eliminates dynamic decision-making unpredictability, ensuring consistent system behavior 
across executions and simplifying debugging and performance analysis.

\stitle{Limitations of static scheduling}.
While static scheduling provides strong guarantees and predictable behavior, it has limitations 
affecting performance and flexibility. For highly variable workloads, static scheduling cannot adapt 
to changing execution patterns, leading to suboptimal resource utilization. This is especially problematic 
when task execution times vary significantly with input characteristics or have dynamic control flows. 
Static scheduling's effectiveness also depends on accurate offline profiling, which can be compromised 
by concurrent task interference, hardware state variations (e.g., thermal throttling), and input data characteristics, 
potentially leading to suboptimal XPU assignments.

\stitle{Dynamic scheduling extensions for {\sys}}.
We discuss several extensions that can be added to {\sys} to support dynamic scheduling.
We leave this as future work.

\etitle{Dynamic priority adjustment.}
{\sys} natively supports dynamic priorities, 
as priority information can be updated before each {\xnode} invocation (see $\S$\ref{sec:sched}). 
This capability enables adaptation to dynamic workloads, 
such as responding to critical events by elevating associated task priorities.

\etitle{Dynamic XPU assignment.}
Currently, {\sys} does not support runtime XPU reassignment.
However, the {\xnode} abstraction in {\sys} makes dynamic XPU selection theoretically feasible, 
as XPU selection is managed entirely by the runtime rather than requiring user intervention. 
Implementing such a feature would require addressing several challenges, including 
efficient data migration between XPUs and maintaining consistency for stateful operations.

\etitle{Dynamic scheduling policies.}
Our ILP-based approach could serve as a valuable foundation for dynamic scheduling policies.
The system could pre-compute multiple scheduling solutions using the ILP formulation for different
anticipated execution time scenarios and dynamically select the appropriate one based on runtime observations.
Alternatively, simpler heuristic policies could be employed for runtime adaptation,
with the ILP-based solution serving as a fallback.

\section{Related Work}\label{sec:related}

\nostitle{Systems for AuT applications.}
ROS~\cite{ROSPaper,ROSWebsite} has emerged as the predominant framework for 
AuT applications~\cite{Smolyanskiy2017TowardLA}.
However, vanilla ROS lacks real-time support and native task scheduling~\cite{Saito2018ROSCHRealTimeSF,
Saito2016PriorityAS,Wei2016RTROSAR}.
Subsequently, ROS2~\cite{ROS2Paper,NextROS} introduced real-time scheduling
for CPU tasks via Executor concept~\cite{Yang2020ExploringRE,Jiang2022RealTimeSA}.
Similarly, CyberRT~\cite{CyberRT}, developed for Apollo autonomous driving platform~\cite{Apollo},
supports user-level CPU task scheduling using coroutines.
ERDOS~\cite{Gog2022D3AD}, a state-of-the-art middleware for autonomous driving
applications, advances these capabilities by supporting dynamic deadlines and
algorithm implementations for each module.
However, these frameworks are limited to scheduling CPU-based module-level tasks
and do not address GPU or NPU operations.
The NVIDIA Compute Graph Framework (NVCGF)~\cite{nvcgf}, a commercial framework
for autonomous driving, stands alone in providing
fine-grained task scheduling for NVIDIA GPUs and DLAs. 
We didn't include NVCGF in our evaluation because it only supports the 
commercial Drive Orin platform (not Jetson Orin platform).
NVCGF requires developers
to partition tasks into passes, each executed on a specific XPU, similar to our
stage-level {\xnode} abstraction. It implements pass scheduling based on graph
topology and WCET of each pass.
{\sys} is more flexible than NVCGF in two key aspects: {\sys} supports multi-XPU implementations
for each stage, and enables preemptive XPU scheduling, whereas NVCGF is limited to 
non-preemptive scheduling.
Several academic studies~\cite{Enright2024PAAMAF,Amarnath2021HeterogeneityAwareSO,
Alcaide2019HighIntegrityGD} have proposed improvements to GPU task scheduling in
AuT applications, our work distinguishes itself by addressing both XPU assignment
and scheduling holistically.

\stitle{Systems for cooperating XPUs.}
The coordination of multiple XPUs has been studied in many domains.
For example, Molecule~\cite{Du2022Serverless} offload serverless functions to
DPU and FPGAs.
Band~\cite{Jeong2022BandCM} and a prior work~\cite{Seo2021SLOAwareIS} propose
DNN inference serving systems on edge and mobile platforms that utilizes CPU, GPU and NPU.
PolyMath~\cite{Kinzer2021ACS} and Yin-Yang~\cite{Kim2022YinYangPA} 
propose high-level programming abstractions that can be mapped 
to cross-domain accelerators. 
Different from these works, {\sys} focuses on
both the task scheduling and XPU assignment for AuT applications.
Different domains face distinct challenges when coordinating multiple XPUs.
For DNN inference serving systems, the primary challenge lies in load balancing
multiple inference requests across XPUs while meeting strict
Service Level Objectives (SLOs)~\cite{Seo2021SLOAwareIS}. In contrast, for
AuT applications, the key challenge is holistically optimizing
tens of algorithm modules with complex topological relationships.

\stitle{XPU tasks scheduling.}
Due to the increasing popularity of heterogeneous computing, plentiful researches have
been proposed to schedule tasks for different XPUs, including preemptive scheduling for
GPUs~\cite{Wu2017FLEPEF,Chen2017EffiShaAS,han2022reef,Calhoun2012PreemptionOA,Kato2011TimeGraphGS,Zhou2015GPESAP,CTXBack,Lin2016EnablingEP,Capodieci2018DeadlineBasedSF,Wang2024GCAPSGC}
and NPUs~\cite{Choi2019PREMAAP,Kouris2022FluidBE},
concurrency enhancement on
GPUs~\cite{Wang2010KernelFA,Jain2019FractionalGS,Cui2021EnableSD,Liang2015EfficientGS,Chen2016BaymaxQA,Chen2017ProphetPQ,Wang2016SimultaneousMG},
TPUs~\cite{Baek2020AMN,Yin2022ExactMA},
and NPUs~\cite{Xue2023V10HN,Lee2021DataflowMA},
and virtualization of
GPUs~\cite{Gu2018GaiaGPUSG} and FPGAs~\cite{Landgraf2021CompilerdrivenFV}.
While these works have made significant contributions to XPU scheduling, they often focus on specific hardware types. 
Building upon these advances, {\sys} focuses on enabling preemptive scheduling for each XPU in multi-XPU applications. 
Specifically, {\sys} leverages the same mechanism of the GPU driver, as used in a recent work~\cite{Wang2024GCAPSGC}.
On the other hand, {\sys} is orthogonal to prior works. Many of these works can be integrated 
into {\sys}, e.g., reset-based preemption on GPUs~\cite{han2022reef,Park2015ChimeraCP} and NPUs~\cite{Choi2019PREMAAP},
to implement the two APIs \textit{preempt} and \textit{restore}.

\stitle{DAG Scheduling.}
DAG-like programming models are widely used in many domains, such as
big data processing~\cite{Carbone2015ApacheF,Zhuang2023EXOFLOW,MapReduceOSDI},
machine learning~\cite{Abadi2016TensorFlow,PaLMPathWays},
and cyber physical systems~\cite{Cong2010SupportingSR}.
Therefore, many algorithms have been proposed for scheduling DAGs on
multiprocessor systems~\cite{Baruah2020SchedulingDW,
Dai2022ResponseTimeAO,Guan2021DAGFluidAR,Baruah2015TheFS,Baruah2015TheGE}.
For instance, Baruah~\cite{Baruah2020SchedulingDW} proposed an ILP model to construct
a time table that minimizes end-to-end latency for a DAG, 
where each node has already been assigned to a specific CPU. 
While their approach focuses on CPU scheduling with fixed assignments, our work extends this 
concept to include additional variables such as XPU type selection, XPU assignment, and task priorities.

\section{Conclusion}\label{sec:concl}
This paper proposes {\xnode}, a fine-grained multi-XPU abstraction that
enables stage-level XPU-aware task management. This abstraction simplifies development by 
allowing programmers to focus on XPU-specific implementations while handling XPU selection 
and coordination automatically. Based on this abstraction, we developed {\sys}, a runtime system
that provides comprehensive XPU assignment and scheduling capabilities.
Experiments on two heterogeneous platforms demonstrate the 
effectiveness of {\sys}, and emphasizes the importance of holistic XPU management.

\small{
\bibliographystyle{plain} 
\bibliography{xdag}
}

\end{document}